\documentclass{aastex631}
\usepackage{hyperref}
\def\apjs{{\rm ApJS}}                  % Astrophys. J. Supl.
\def\apjl{{\rm ApJ Let}}               % Astrophys. J. (Letters)
\def\Msol{\thinspace\hbox{$\hbox{M}_{\odot}$}}

\def\a4{\hsize 17.0cm \vsize 25.cm}
\newcommand{\der}[2]  { \frac{{\rm d}#1}{{\rm d}#2} }

\shorttitle{Molecular gas in young stellar clusters}
\shortauthors{Silich  et al.}

\begin{document}

\title{Molecular gas properties in young stellar clusters with a suppressed
       star cluster wind}

\author{Sergiy Silich}
\affiliation{Instituto Nacional de Astrof\'\i sica \'Optica y Electr\'onica,
AP 51, 72000, Puebla, M\'exico \\ email: silich@inaoep.mx}
\author{Jean Turner}
\affiliation{UCLA Department of Physics and Astronomy,
             CA 90095-1547, Los Angeles, USA}
\author{Jonathan Mackey}
\affiliation{Dublin Institute for Advanced Studies, Astronomy \&
Astrophysics Section, DIAS Dunsink Observatory, Dublin \\
D15 XR2R, Ireland}
\author{Sergio Mart\'{\i}nez-Gonz\'alez}
\affiliation{Instituto Nacional de Astrof\'\i sica \'Optica y Electr\'onica,
AP 51, 72000, Puebla, M\'exico}

\date{Accepted . Received ; in original form}

%\pubyear{2021}

\begin{abstract}
In compact and dense star-forming clouds a global star cluster wind
could be suppressed. In this case the stellar feedback is unable to expel the
leftover gas from the cluster. Young massive stars remain embedded into a dense
residual gas and stir it moving in the gravitational well of the system. Here
we present a self-consistent model for the molecular gas distribution
in such young, enshrouded stellar clusters. It is assumed that the cloud
collapse terminates and the star formation ceases when a balance between the
turbulent pressure and gravity and between the turbulent energy dissipation
and regeneration rates is established. These conditions result in an equation
that determines the residual gas density distribution that, in turn, allows
one to determine the other characteristics of the leftover gas and the star
formation efficiency. It is shown that model predictions are in good
agreement with several observationally determined properties of cloud
D1 in nearby dwarf spheroidal galaxy NGC 5253 and its embedded cluster.

\end{abstract}


\section{Introduction}
\label{sec:1}

Star formation is a fundamental problem in astrophysics for many years.
There is a general agreement that star formation requires the dense
molecular gas and occurs in giant molecular clouds. It was shown long
time ago that the interstellar gas depletion time in our Galaxy is
much longer than the free-fall time of molecular clouds
\citep[e.g.][]{Williams1997}
that raised the problem of the molecular cloud stability
and apparently low average star formation efficiency in our and other
galaxies. Different solutions to these problems have been proposed and
include strong magnetic fields \citep[e.g.][]{Shu1987},
photoionization of the intracloud gas \citep{McKee1989,Franco1994}
and different modifications of the feedback scenario 
- the injection of energy and momentum into the residual molecular gas by
low-mass \citep{Norman1980,McKee1989} or high-mass \citep{Matzner2002} stars
\citep[see a review by][]{Padoan2014}.

In contrast, to form a bound stellar cluster a large ($>$ 30\%) star
formation efficiency is required \citep{Baumgardt2007, Baumgardt2008}.
This raises the question how to prevent or at least retard the leftover gas
expulsion and the molecular cloud disruption at early stages of massive star
cluster formation (e.g. \cite{Wirth2022} claimed that it took 3.5Myr to
4.0Myr to stop star formation in Galactic globular clusters).

The supersonic velocity dispersion has been detected in objects of different
scales from cores of molecular clouds and compact, young, still enshrouded by
molecular gas stellar clusters, to giant HII regions and HII galaxies. This
led \cite{Terlevich1981}, \cite{Solomon1987} and \cite{Melnick1987} to suggest
that supersonic turbulence is a characteristic of the gas virial equilibrium 
and that the dynamic feedback from the newborn stars prevents the parental
cloud from the further collapse.

\cite{TenorioTagle1993} associated supersonic velocities with a collection
of bow shocks around pre-main-sequence stars moving in the gravitational well
of the cluster and suggested that these bow shocks stir and maintain  
supersonic turbulence until massive stars and supernovae expel the remaining
gas from the cluster. However, the supersonic turbulence decays very rapidly
\citep{Stone1998,MacLow1999} that requires sources more powerful than
low-mass stars to maintain supersonic turbulence in star-forming clouds
\citep[e.g.][]{Murray2010}. Different aspects of the supersonic turbulence
in molecular clouds were reviewed by, among others, \cite{Scalo1987},
\cite{Vazquez2000}, \cite{Elmegreen2004}, \cite{MacLow2004}, \cite{McKee2007},
\cite{Padoan2014}. Bow shocks, their turbulent mixing layers and wakes were also
discussed by many authors \citep[see][and references therein]{Arthur2006,
Wareing2007,Binette2009,Mackey2013,Mackey2015,Henney2019}.

The role of massive stars in the turbulent energy regeneration was discussed by
\cite{Matzner2002} who considered HII regions as the major mechanism for the
turbulent energy regeneration, but did not discuss the effects of the
intracluster gas distribution.

\cite{Marks2012} found that initial star cluster half-mass radii
weakly depend on stellar mass (see their equation 7) and even for
$10^5 - 10^6$\Msol \, clusters hardly exceed 1pc. In such compact and massive
star-forming regions wind-driven bubbles around individual massive stars
stall before merging with their neighbors \citep[see][]{Silich2017, Silich2018}
and ionizing photons are effectively absorbed by dust grains which re-emit
them in the infrared waveband. Upon such conditions massive stars are not able
to photoionize the bulk of leftover gas, form a global star cluster wind and
expel the residual gas from the star-forming region. The negative stellar
feedback is then suppressed. Instead, ultracompact HII regions (UCHII)
embedded into the residual molecular gas are formed \citep{Silich2020}.
However, \cite{Silich2017, Silich2018} considered only the mechanical
equilibrium in star-forming clouds with an arbitrary selected star formation
efficiency (SFE) and did not account for the rapid turbulent energy
dissipation.

Here, following \cite{Norman1980, McKee1989, Matzner2002} we assume
that the pre-stellar cloud collapse is followed by a vigorous star
formation that terminates when a balance between the turbulent pressure and
gravity, and the turbulent energy dissipation and regeneration rates is
established. In contrast with other authors \citep[e.g.][]{Padoan1995,
Krumholz2005,Krumholz2006,Padoan2011}, we do not consider how star formation
proceeds in the collapsing cloud, but assume that it results in a stellar
cluster with a certain mass and 
a known stellar density distribution and that the further star formation is
altered by the stellar feedback. It is also postulated that the cluster is
sufficiently compact and dense to prevent the leftover gas expulsion,
the post-star-forming system is dynamically stable and that stellar feedback
compensates the turbulent energy dissipation continuously. Our
aim is to find out how the leftover molecular gas is distributed and obtain
its other characteristics upon such conditions. We show that
the equilibrium conditions together with the only one free parameter that
characterises the degree to which the feedback energy could be conserved, 
fix the residual gas density, velocity dispersion and temperature
distributions and allow one to estimate the star formation efficiency
if the stellar mass distribution is known.

The paper is organized as follows. In section 2 we select a model for the
stellar mass distribution. In section 3 the conditions for the thermal and
mechanical equilibrium are formulated and the equation that determines the
residual gas density distribution is derived. We demonstrate then that the
velocity dispersion and the star formation efficiency in the post-star-forming
cloud follow directly from the equilibrium conditions and the gas density
distribution. In section 3.3 we discuss the major
sources for the molecular gas heating and cooling and show how to determine
the molecular gas temperature. In section 4 we confront our model to the well
studied molecular cloud D1 in the nearby dwarf spheroidal galaxy NGC 5253 and
show that model predictions are in agreement with several observational
characteristics of this cloud and its embedded cluster. Finally, in section 5
we summarize our findings and the major model restrictions.

\section{Stellar mass distribution}

Hereafter it is assumed that star formation in the parental molecular cloud
results in a dense compact cluster with a total stellar mass $M_{SC}$ and a
Gaussian stellar density distribution:
%---------------------------------------------------------------
\begin{eqnarray}
      \label{eq1a}
      & & \hspace{-0.9cm} 
\rho_{\star}(r) =  \frac{M_{SC}}{(2\pi)^{3/2} b^3} \exp{\left[-\frac{1}{2}
                   \left(\frac{r}{b}\right)^2\right]} ,
      \\[0.2cm] \label{eq1b}
      & & \hspace{-0.9cm} 
n_{\star}(r) =  \frac{N_{\star}}{(2\pi)^{3/2} b^3} \exp{\left[-\frac{1}{2}
                   \left(\frac{r}{b}\right)^2\right]} ,                
\end{eqnarray}
%---------------------------------------------------------------
where $\rho_{\star}$ is the stellar mass density, $N_{\star}$ and $n_{\star}$
are the total number and the number density of the turbulence-driven stars,
respectively. $b$ is the star cluster core radius. 

The stellar mass $M_{\star}(r)$ enclosed within a sphere of radius $r$ then
is: 
%---------------------------------------------------------------
\begin{equation}
\label{eq2}
M_{\star}(r) = M_{SC} \left[erf \left(\frac{r}{2^{1/2} b}\right) - 
       \left(\frac{2}{\pi}\right)^{1/2} \frac{r}{b}
               \exp\left[-{\frac{1}{2}
                   \left(\frac{r}{b}\right)^2}\right]\right] ,
\end{equation}                   
%---------------------------------------------------------------
where $erf(r)$ is the error function.  

\section{Equilibrium conditions}

\subsection{Turbulent energy dissipation and regeneration rates}

Following \cite{Stone1998, MacLow1999, Basu2001} we assume that the rate of the
turbulent energy dissipation in the residual gas is:
%---------------------------------------------------------------
\begin{equation}
\label{eq3}
Q_{dis}(r) = {\frac{\eta_d \rho_g \sigma^3}{\lambda}} , 
\end{equation}
%---------------------------------------------------------------
where $\rho_g$ and $\sigma$ are the residual gas density and 1D velocity
dispersion, respectively, and $\lambda$ is the turbulence driving
scale. The dimensionless factor $\eta_d \sim 1$ over a range of driving
lengths \citep[see][and references therein]{Basu2001}. We assume that
$\eta_d = 1$ in all our simulations.

In spite of many discussions \citep[e.g.][]{Basu2001,Quillen2005,Swift2008,
Brunt2009} the nature
and the value of the driving length remain uncertain. Here we postulate
that turbulence in a newborn cluster is supported by massive stars that
randomly move in the gravitational well of the cluster. It is likely
that in such a case the driving length $\lambda$ is determined by the 
separation between neighboring massive stars. The plausible assumption
then is that the half of the driving wavelength is equal to the mean distance
between the two neighboring massive stars and thus $\lambda = 4 X$, where
$X$, the half-distance between neighboring massive stars, is
\citep[see][]{Silich2020}:
%---------------------------------------------------------------
\begin{equation}
\label{eq4}
X(r) = b \left[(9 \pi / 2)^{1/2}
         \exp{\left[\frac{1}{2}\left(\frac{r}{b}\right)^2\right]}\right]^{1/3}
           N_{\star}^{-1/3} .
\end{equation}
%--------------------------------------------------------------- 
Note that in this case the driving length is not constant as it is adopted
in most models, but depends on the distance from the star cluster
center \citep[the multi-scale energy injection was discussed by][]{Scalo1987}.
This assumption results in the turbulent energy dissipation rate
%---------------------------------------------------------------
\begin{equation}
\label{eq5}
Q_{dis}(r) = \frac{\eta_d \rho_g \sigma^3 N^{1/3}_{\star}}
            {4 (9 \pi/ 2)^{1/6} b 
             \exp\left[\frac{1}{6} (\frac{r}{b})^2\right]} . 
\end{equation}
%---------------------------------------------------------------

We assume that the lost turbulent energy is regenerated in the turbulent
mixing layers and turbulent wakes formed around most massive stars that
move in the gravitational well of the cluster. We further follow arguments
presented in \cite{Norman1980,McKee1989,Matzner2002} and assume that recently
formed stars support turbulence in the residual gas at the rate
%---------------------------------------------------------------
\begin{equation}
\label{eq6}
Q_{\star}(r) = P \sigma(r) n_{\star}(r) / 2, 
\end{equation}
%---------------------------------------------------------------
where $P$ is the average rate of momentum input to the intracloud
medium per massive star that includes the momentum input via the stellar
wind ${\dot M_w} V_w$, radiation $L_{\star}/c$ and that, due to the
photo-heated HII region expansion.
The precise value of the momentum continuously returned to the intracloud
medium by a typical massive star is uncertain. Here we normalize it to the
stellar radiative momentum input rate \citep[see][]{Henney2019}:
%---------------------------------------------------------------
\begin{equation}
\label{eq7}
P = \eta_{\star} L_{\star} / c , 
\end{equation}
%---------------------------------------------------------------
where $L_{\star} = L_{bol} / N_{\star}$ is the average turbulence-driving star
luminosity, $L_{bol}$ is the star cluster bolometric luminosity and $c$ is
the speed of light. The
$\eta_{\star}$ factor is considered as a free parameter of the model. The value
of this parameter is determined by the degree to which the feedback energy
could be conserved. It depends on the physical conditions in the star-forming
cloud, and in general must be determined by numerical simulations.
\cite{McKee1999}  suggested that $\eta_{\star} \approx 1.6$, to account for the
energy stored in the magnetic field, while numerical simulations by Mackey
(2013), focused on the dynamics of HII regions around moving O stars, showed
that the momentum input rate could exceed that provided by radiation
pressure up to 10 times. It is interesting to note that AGN outflows also
often have momentum well in excess (up to 30 times) of the central black hole
integrated momentum $L_{BH} \tau_{in} /c $, where $\tau_{in}$ is the
characteristic time scale of the radiative feedback \citep[see][and
references therein]{Faucher2012}. We adopt that
$\eta_{\star}$ falls in the range $1 < \eta_{\star} < 10$.

In the dense, compact clusters the characteristic size of individual
wind-driven bubbles are much smaller than the cluster core radius
$b$ \citep{Silich2017,Silich2018}. Therefore we assume that the energy
dissipation and regeneration rates are balanced locally throughout the
cluster: $Q_{dis}(r) = Q_{\star}(r)$. This leads to the relation:
%---------------------------------------------------------------
\begin{equation}
\label{eq8}
\rho_g \sigma^2 = \frac{\eta_{\star}}{\eta_d} \left(\frac{3}{4\pi^4}\right)^{1/3}
                  \frac{L_{bol}}{c b^2 N^{1/3}_{\star}} 
                  \exp{\left[-\frac{1}{3} \left(\frac{r}{b}\right)^2\right]} . 
\end{equation}
%---------------------------------------------------------------

\subsection{Mechanical equilibrium}

The mechanical equilibrium in the post-star-forming cloud requires 
the gravitational pull of the cluster to be in balance with the
turbulent pressure \citep[e.g.][]{Calura2015}
%---------------------------------------------------------------
\begin{equation}
\label{eq9}
\der{P_{turb}}{r} = - \frac{G \rho_g (M_g(r) + M_{\star}(r))}{r^2} , 
\end{equation}
%---------------------------------------------------------------
where $G$ is the gravitational constant, $M_g(r)$ and $M_{\star}(r)$ are
the gas and the stellar mass enclosed within a sphere of radius $r$,
respectively, $P_{turb}(r) = \rho_g \sigma^2$ is the residual gas
turbulent pressure.

\subsection{Equilibrium gas distribution}

Equation (\ref{eq8}) allows one to calculate the
turbulent pressure derivative: ${\rm d}{P_{turb}}/{\rm d}{r} =
{\rm d}(\rho_g \sigma^2)/{\rm d}{r}$. Combining this derivative with equation
(\ref{eq9}), one can obtain an equation that determines the residual gas
density distribution in a stationary post-star-forming cloud upon the
assumption that the turbulent energy dissipation is continuously compensated
by the stellar feedback:
%---------------------------------------------------------------
\begin{eqnarray}
      \nonumber
      & & \hspace{-0.9cm} 
\rho_g(r) = \frac{\eta_{\star}}{\eta_d} \left(\frac{2}{9 \pi}\right)^{1/3}
            \frac{L_{bol}}{\pi c G b N^{1/3}_{\star} [M_{\star}(r) + M_g(r)]}
            \left(\frac{r}{b}\right)^3
            \exp\left[-\frac{1}{3} \left(\frac{r}{b}\right)^2\right]
      \\[0.2cm] \label{eq10}
      & & \hspace{-0.9cm}
      \approx 10^{-14}  \frac{\eta_{\star}}{\eta_d}
            \frac{L_{42}}{N^{1/3}_{\star} b_1 [M_{\star,s}(r) + M_{g,s}(r)]}
            \left(\frac{r}{b}\right)^3
            \exp\left[-\frac{1}{3} \left(\frac{r}{b}\right)^2\right]
            \quad g \, \, cm^{-3} ,
\end{eqnarray}
%---------------------------------------------------------------
where $L_{42} = L_{bol} / 10^{42}$erg s$^{-1}$, $b_1$ is the core radius
in pc units, $M_{\star,s}(r)$ is the stellar and $M_{g,s}(r)$ is the gas mass
in solar units. The residual gas mass in equation (\ref{eq10}) is: 
%---------------------------------------------------------------
\begin{equation}
\label{eq11}
M_g(r) = 4 \pi \int_0^r x^2 \rho_g(x) {\rm d} x . 
\end{equation}
%---------------------------------------------------------------
We solve equation (\ref{eq10}) by iterations by making use equation
(\ref{eq11}). At the first step it is assumed that the gas density is equal
to zero in the whole star cluster volume. This implies that in equation 
(\ref{eq10}) $M_g(r) = 0$, but allows one to calculate a new, nonzero gas
density distribution $\rho_i(r)$ as the stellar density and stellar mass
are not equal to zero. A new gas density distribution is used then to integrate
equation (\ref{eq11}) numerically and obtain a new, nonzero gas mass
distribution $M_g(r)$. This $M_g(r)$ is used in equation (\ref{eq10}) to
improve the gas density distribution. The iteration process continues until
the difference between the subsequent values of the integrated gas mass
becomes small enough: 
$\mid(M_{g, i} - M_{g, i-1})\mid / (M_{g, i} + M_{g, i-1}) < \epsilon$.
We usually stop iterations when $\epsilon$ drops below $10^{-5}$.
After the residual gas distribution was calculated, one can easily obtain
the other characteristics of the star-forming region. The velocity
dispersion is calculated by means of equations (\ref{eq8}) and (\ref{eq10}):
%---------------------------------------------------------------
\begin{eqnarray}
      \nonumber
      & & \hspace{-0.9cm}
      \sigma(r) = \left[\frac{3 G (M_{\star}(r) + M_g(r))}{2 b}\right]^{1/2}
      \left(\frac{b}{r}\right)^{3/2}
      \\[0.2cm] \label{eq12}
      & & \hspace{-0.9cm}
      \approx 8 \times 10^{-2}
      \left[\frac{M_{\star,s}(r) + M_{g,s}(r)}{b_1}\right]^{1/2}
      \left(\frac{b}{r}\right)^{3/2} \quad km \, \, s^{-1} .
\end{eqnarray}
%---------------------------------------------------------------
The star formation efficiency is:
%---------------------------------------------------------------
\begin{equation}
\label{eq13}
SFE = M_{SC} / [M_{SC} + M_g(R)] ,
\end{equation}
%---------------------------------------------------------------
where $R$ is the adopted radius of the cluster. The total turbulent
energy dissipation rate in the residual gas is:
%---------------------------------------------------------------
\begin{equation}
\label{eq14}
L_{gas}(R) = 4 \pi \int_0^R {Q_{dis}(x) x^2 {\rm d} x} .
\end{equation}
%---------------------------------------------------------------

The stellar bolometric and the wind mechanical luminosities do not increase
linearly with the stellar mass. We estimate that only (15\% - 20\%) of
massive stars contribute about 90\% to the star cluster energy budget and
adopt for the number of the turbulence-driven stars
%---------------------------------------------------------------
\begin{equation}
      \label{eq16}
N_{\star} \approx 0.15 \times N_{massive} (M_{SC}/10^6\Msol) ,
\end{equation}
%---------------------------------------------------------------
where $N_{massive}$ is the number of massive ($M > 8$\Msol) stars in a
$10^6$\Msol \, cluster. In clusters with a canonical Kroupa IMF
$N_{massive} \approx 1.1 \times 10^4$ (e.g. Calura 2015), while in the case of
a Salpeter IMF with the lower and upper cutoff masses 3\Msol \, and
120\Msol, respectively, $N_{massive} \approx 3.1 \times 10^4$.
 
\subsection{Thermal balance and the molecular gas temperature}

The thermal balance in a post-star-forming cloud is determined by the turbulent,
cosmic ray and X-ray gas heating and the molecular gas cooling rates
\citep[e.g.][]{Maloney1996,Basu2001,Shang2002,Pan2009,Papadopoulos2010}:
%---------------------------------------------------------------
\begin{equation}
      \label{eq17}
Q_{dis} + Q_{CR} + Q_{XR} - Q_{cool} - Q_{gd} = 0 .
\end{equation}
%---------------------------------------------------------------
In equation (\ref{eq17}) $Q_{dis}$ is the turbulent energy dissipation rate
(see equation \ref{eq3}), $Q_{CR}$ and $Q_{XR}$ are the cosmic ray and the X-ray
heating rates and $Q_{cool}$ is the molecular gas cooling rate, respectively.
$Q_{gd}$ is the energy exchange between the gas and dust grains
\citep[see][]{Goldsmith2001,Pan2009,Papadopoulos2010}:
%---------------------------------------------------------------
\begin{equation}
      \label{eq18}
Q_{gd} = 7 \times 10^{-34} n^2 T_g^{1/2} (T_g - T_d) \, erg \, cm^{-3} \, s^{-1} ,
\end{equation}
%---------------------------------------------------------------
where $n = \rho_g / \mu_{mol}$ is the molecular gas number density,
$\mu_{mol}$ is the mean mass per particle in the molecular gas, 
$T_g$ and $T_d$ are the molecular gas and the dust grain temperatures.
We adopt $\mu_{mol} = 2.33 m_H$, where $m_H$ is the hydrogen atom mass. 
In the UV-shielded dense environment CR heating rate is determined by the
CR ionization rate per H$_2$ molecule \cite{Papadopoulos2010}:
%---------------------------------------------------------------
\begin{equation}
      \label{eq19}
Q_{CR} = 1.5 \times 10^{-24} \xi_{CR,17} \, n_4 \, erg \, cm^{-3} \, s^{-1} ,
\end{equation}
%---------------------------------------------------------------
where $\xi_{CR,17}$ is the CR ionization rate in $10^{-17}$ s$^{-1}$ 
and $n_4$ is the molecular gas number density in the $10^4$ cm$^{-3}$ units.
The reference value for the $\xi_{CR}$ is $5 \times 10^{-17}$ s$^{-1}$ -
the average CR ionization rate in our Galaxy. In compact starbursts it
may be up to $10^3$ times larger \cite{Papadopoulos2010}.

The residual molecular gas in dense UV-shielded parcels of the star-forming
cloud may be also exposed to and heated by the soft and hard X-ray
emission caused by the magnetic lines reconnection in young stellar
objects, X-ray binaries or shock-heated stellar winds
\citep[e.g.][and references therein]{Maloney1996,Feigelson1999,Shang2002,
Meijering2005}. For example, \cite{Tsujimoto2006} reported the detection of
hard X-ray emission from the two UCHIIs in W49A, one of the most active
star-forming regions in our Galaxy. The X-ray heating rate is given by
\citep{Panoglou2012,Mackey2019}:
%---------------------------------------------------------------
\begin{equation}
      \label{eq20}
Q_{XR} = \eta_X n H_X ,
\end{equation}
%----------------------------------------------------------------
where $\eta_X$ is the heating efficiency \citep[see][]{Dalgarno1999}.
$H_X$ is the X-ray energy absorption rate per particle
\citep{Maloney1996,Panoglou2012}:
%---------------------------------------------------------------
\begin{equation}
      \label{eq21}
H_X = \int_{E_{min}}^{E_{max}} \sigma_X(E) \, F(E) \, {\rm d} E ,
\end{equation}
%---------------------------------------------------------------
where $F(E)$ is the X-ray flux and $\sigma_X(E)$ is the photoelectric
cross-section per H nucleus.

One can obtain the molecular gas temperature from the energy balance
equation (\ref{eq17}) by making use a reasonable approximation for the
molecular gas cooling rate $Q_{cool} = Q_{cool}(n,T,{\rm d}u/{\rm d}r)$,
where ${\rm d}u/{\rm d}r$ is the velocity gradient.

\section{NGC 5253 cloud D1 and its embedded cluster}

In this section we confront our model with NGC 5253 D1 molecular cloud and
its young, compact cluster. For the star cluster mass, core radius and
metallicity we adopt $M_{SC} = 1.625 \times 10^5$\Msol , $b = 0.8$~pc and
$Z = 0.2 Z_{\odot}$ \citep[see][]{Turner2017,Silich2020}.  We also adopt a
canonical Kroupa initial mass function with lower and upper cutoffs
$M_{low} = 0.1$\Msol \, and $M_{up} = 120$\Msol, respectively.
In this case $N_{\star} \approx 270$. We then make use Starburst99 synthetic
model to obtain the embedded cluster bolometric luminosity
($L_{bol} = 3.9 \times 10^{42}$~erg s$^{-1}$  at the age of 1 Myr)
and calculate the residual gas density distribution. The velocity dispersion
is then obtained from equation (\ref{eq12}). The stellar mass and the
model-predicted gas density distributions are presented by dotted, solid and
dashed lines on Fig. 1 panel a, while the velocity
dispersion is shown in panel b of this figure. The dotted line on panel a
displays the stellar density distribution derived from the Gaussian fit to the
radio and infrared (IR) integrated intensity maps
\citep[see][]{Turner2000,Gorjian2001,Turner2015,Turner2017}, while the solid
and dashed lines present the model-predicted molecular gas distribution for
$\eta_{\star} = 1$ and $\eta_{\star} = 5$, respectively. The model predicts
different stellar and gas density distributions with more extended gas
distribution and stellar mass concentrated towards the star cluster center.
This is consistent with different core radii in Gaussian fits to the observed
IR/radio and molecular gas emissions \citep{Turner2017}.

The corresponding 1D velocity dispersion is shown on panel b.
%----------------------------------------------------------------
\begin{figure}[htbp]
\vspace{8.0cm}
\includegraphics{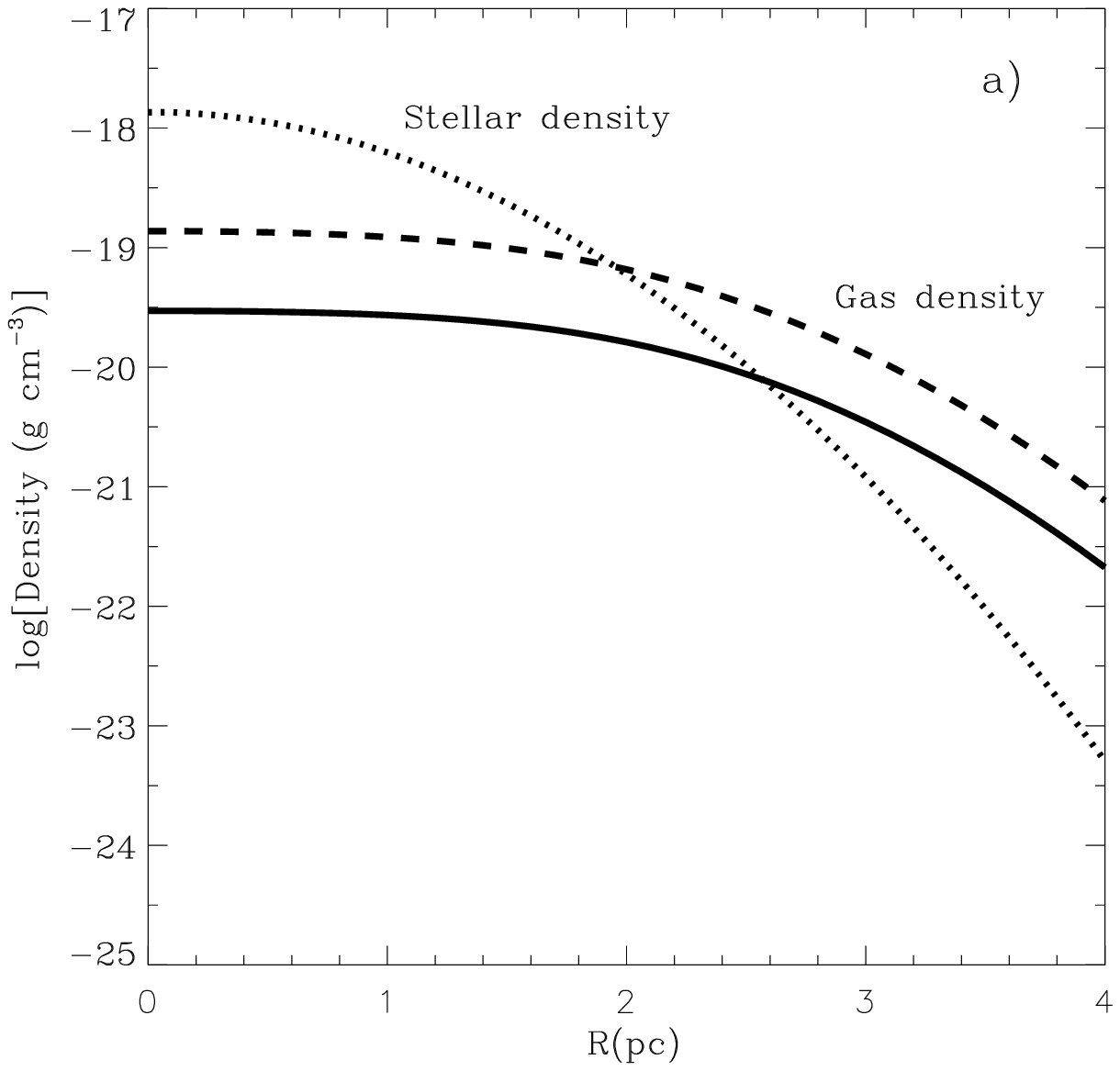}
\includegraphics{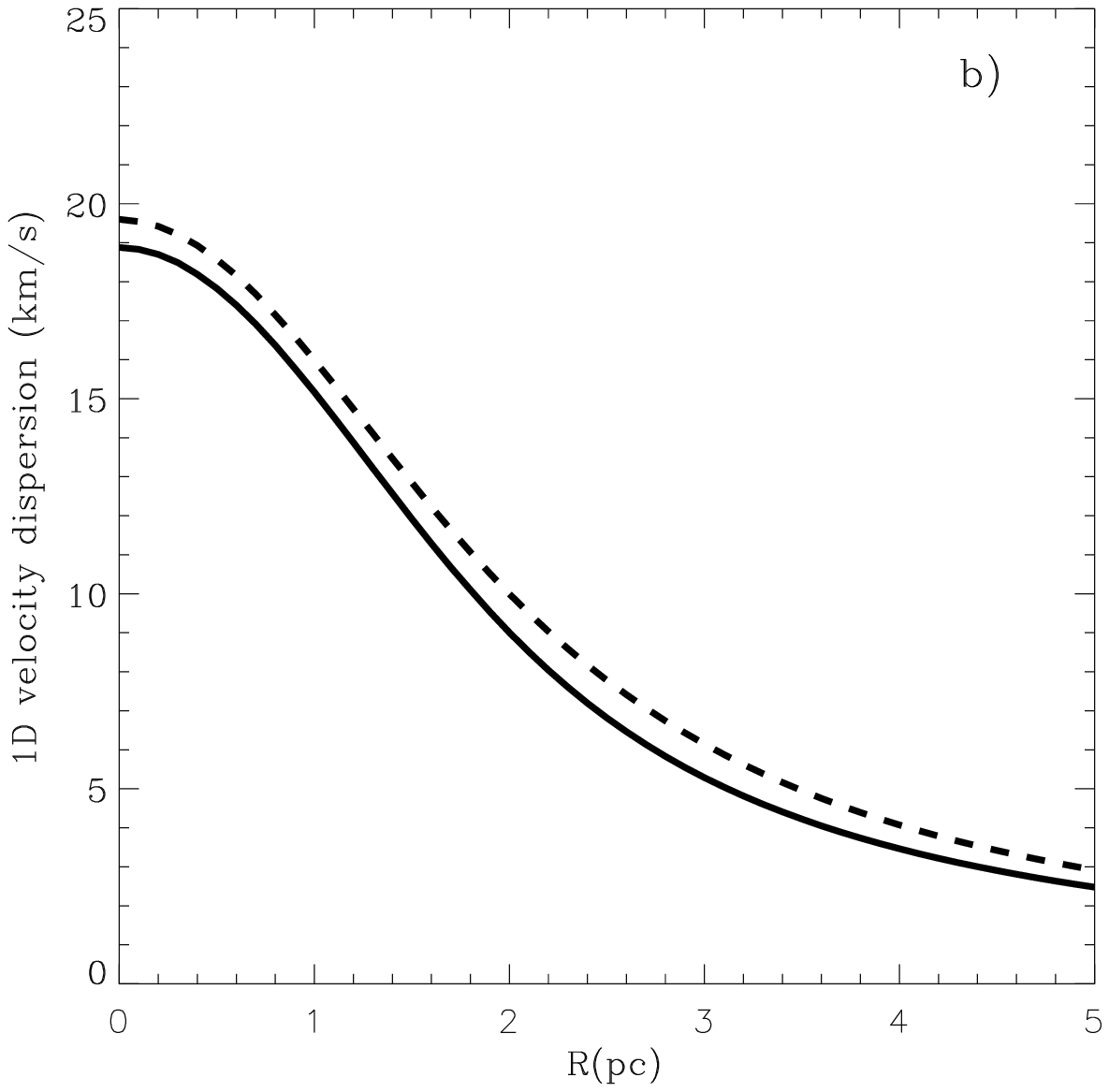}
\caption{The model-predicted D1 cloud structure. The stellar mass
distribution derived from the Gaussian fit to the radio and IR integrated
intensity maps is shown by the dotted line in panel a. The solid and
dashed lines in this panel display the molecular gas distributions obtained
for two different $\eta_{\star}$ factor values: $\eta_{\star} = 1$
(solid line) and $\eta_{\star} = 5$ (dashed line). The model-predicted velocity
dispersion is shown on panel b, where solid and dashed lines correspond
to the same models as in panel a.}
\label{fig1}
\end{figure}
%-----------------------------------------------------------------
Note that the $\eta_{\star}$ value does not affect the velocity dispersion
significantly as $\sigma \sim 1 / \surd{\rho} \sim 1 / \surd{\eta_{\star}}$
(see equation \ref{eq12}). For distant star-forming regions like NGC 5253 D1
cloud, where the available spatial resolution does not allow to study the
velocity dispersion profile, one can make use of the model-predicted velocity
dispersion and gas density distributions to calculate the mass-weighted velocity
dispersion and compare it with the observed value:
%---------------------------------------------------------------
\begin{equation}
      \label{eq24}
\sigma_w = \frac{4 \pi}{ M_g(R)} \int_0^R \rho_g(x) \sigma(x) x^2 {\rm d} x .
\end{equation}
%---------------------------------------------------------------
In the case of D1 cloud the mass-weighted velocity dispersion is
$\sigma_w \approx 9.3$km s$^{-1}$ and $\sigma_w \approx 10.4$km s$^{-1}$ in
models with $\eta_{\star} = 1$ and  $\eta_{\star} = 5$, respectively, that
is in good agreement with the observed CO linewidth \citep[$\sigma \approx
9.2$ km s$^{-1}$][]{Turner2015,Turner2017}. The molecular gas mass within a
7.5~pc radius in these two cases is
$M_g \approx 2.4 \times 10^4$\Msol \, and $M_g \approx 9.5 \times 10^4$\Msol,
respectively, which is consistent with the observed CO emission.

In the model-predicted density range the thermal coupling between the dust
grains and the molecular gas is weak: \cite{Pan2009}, see also Appendix E
in \cite{Whitworth2016}. This implies that
the molecular gas temperature could be evaluated from its own energy balance
and one can neglect the $Q_{gd}$ term in equations (\ref{eq17}) and
(\ref{eq23}). The turbulent heating rate is about
$1.4 \times 10^{37}$erg s$^{-1}$ in the simulations with  $\eta_{\star} = 1$ and
$7.5 \times 10^{37}$erg s$^{-1}$ when  $\eta_{\star} = 5$. In both cases the
integrated turbulent heating exceeds the T Tauri integrated X-ray luminosity
significantly: $L_{XR,TT} = N_{PMS} \times L_{XTT} \approx 1.2 \times 10^{35}$
erg s$^{-1}$, where $N_{PMS} \approx 1.5 \times 10^6 (M_{SC}/10^6$\Msol)
is the number of low mass ($M < 3$\Msol) pre-main sequence stars and
$L_{XTT} \approx 5 \times 10^{29}$erg s$^{-1}$ is a T Tauri star typical
luminosity \citep{Shang2002}. Massive stars emit X-rays at a level
$\sim 10^{-7}$ $L_{bol}$ \citep{Crowther2022} that results in a comparable to
the T Tauri integrated value: $L_{XR,MS} \approx 10^{-7} L_{bol} = 3.9
\times 10^{35}$ erg s$^{-1}$. X-ray emission from the high mass binaries (HMXBs)
may reach ($10^{32} - 10^{33}$) ($M_{SC} / 1 \Msol$) ergs s$^{-1}$
\citep[e.g.][]{MasHesse1999,VanBever2000} and thus be
comparable or even exceed the turbulent heating rate. However, it takes
(4-5)~Myr for the HMXBs to become active. The low mass X-ray binaries (LMXBs)
become active even at later times. In clusters as young as that in the center
of D1 cloud, where the nonthermal radio emission from supernovae was not
detected, X-ray heating by binaries is negligible.

The estimates of the X-ray emission from
hot cometary-like bubbles formed around massive stars with a strong wind
are less certain as depend on the ambient gas density and stellar
parameters. The Chandra observations of UCHII regions 
in Sagittarius B2 \citep{Takagi2002} and W49A \citep{Tsujimoto2006} revealed
hard (3.0keV - 8.0keV) X-ray emission within the range $10^{30}$erg s$^{-1}$ -
$10^{33}$erg s$^{-1}$, associated with some of the UCHII regions. Numerical
modelling of the X-ray emission from the wind-blown bubble around young
moving star BD+60$^o$2522 (the Bubble Nebula) led \cite{Green2019} to the
similar soft X-ray and 1-2 orders of magnitude smaller hard X-ray luminosities.
However, the number of massive stars
with strong stellar winds is much (at least two orders of magnitude)
smaller than that of T Tauri stars. Therefore it is unlikely that the
integrated X-ray heating exceeds the turbulent heating rate unless the
intracluster radiation field is dominated by supermassive stars 
\citep{Smith2016} or the intracluster gas is exposed to the external sources.
Hereafter we neglect gas heating by the X-rays.

The molecular gas cooling rate in this density range could be approximated
by the expression \citep{Ao2013}:
%---------------------------------------------------------------
\begin{equation}
      \label{eq22}
Q_{cool} = 6 \times 10^{-29} n^{1/2} T_g^3 {\rm d}u/{\rm d}r \, erg \, cm^{-3}
\, s^{-1} ,
\end{equation}
%---------------------------------------------------------------
where ${\rm d}u/{\rm d}r$ is the rms velocity gradient  
(${\rm d}u/{\rm d}r = \surd{3} {\rm d}{\sigma}/{\rm d}r$) in units of
km s$^{-1}$ pc$^{-1}$ (see Fig. 2). Note that this approximation must
be taken with some care due to uncertainties in the molecular gas composition,
molecules depletion onto dust grains, emission lines taken into consideration
and their optical depths.
%----------------------------------------------------------------
\begin{figure}[htbp]
\plotone{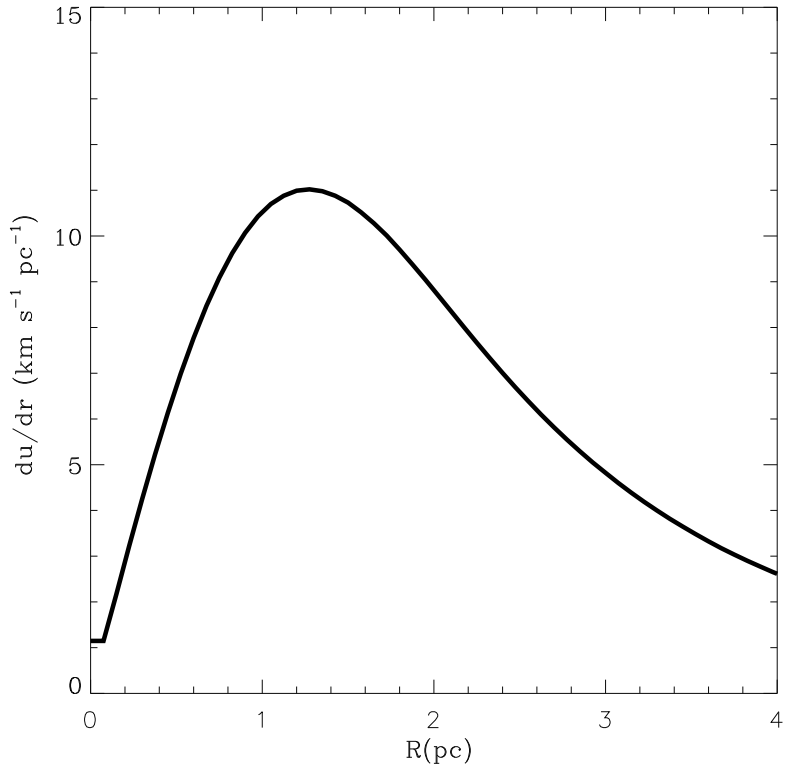}
\caption{The gas velocity gradient calculated upon the assumption that
$\eta_{\star} = 5$.}
\label{fig2}
\end{figure}
%-----------------------------------------------------------------
If X-ray heating can be neglected and the cooling is dominated by
molecular gas whose composition is determined in Table 1 of
\citet{Goldsmith2001}, the energy balance equation (\ref{eq17}) and the
approximation to the gas cooling rate (\ref{eq22}) yield:
%---------------------------------------------------------------
\begin{equation}
      \label{eq23}
T_g = [(Q_{dis} + Q_{CR}) / 6 \times 10^{-29}]^{1/3} ({\rm d}u/{\rm d}r)^{-1/3}
      n^{-1/6} K.
\end{equation}
%---------------------------------------------------------------
The temperature distribution calculated upon the assumption that
$\eta_{\star} = 5$ is shown in Fig. 3. Here the solid, dashed and dotted lines
present the gas temperatures in the cases when CR ionization rate is equal to
that in our Galaxy (see section 3.3), 10 and 100 times larger, respectively.   
%----------------------------------------------------------------
\begin{figure}[htbp]
\plotone{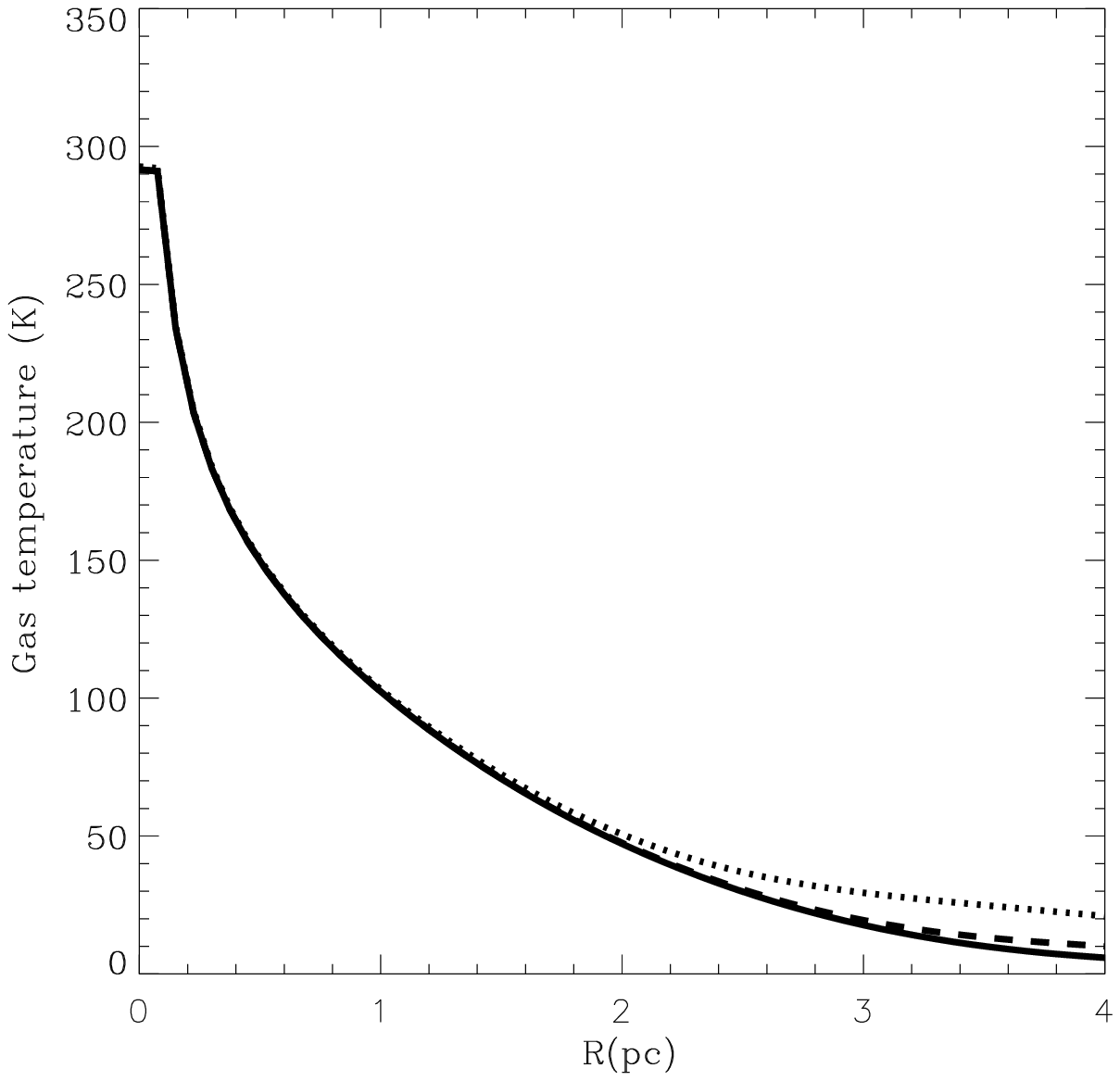}
\caption{The model-predicted gas temperature distribution. Solid, dashed
and dotted lines present the model-predicted temperature distribution in
the cases when the CR ionization rate is equal to that in our Galaxy,
10 and 100 times larger, respectively. The temperatures were calculated
upon the assumption that $\eta_{\star} = 5$.}
\label{fig3}
\end{figure}
%-----------------------------------------------------------------
At the Milky Way CR intensity the contribution of the CR heating is negligible.
However, at starburst-like ionization rates CR heating becomes significant
at the outskirts of the star-forming region as the turbulent heating rate per
unit volume drops with radius fast while it is likely that the CR density
at the D1 cloud scale remains almost homogeneous because of large diffusion
length \citep[e.g.][]{Aharonian2019}. Large predicted
molecular gas temperatures are consistent with the large CO(3-2)
over CO(2-1) intensity ratio in D1 cloud \citep[see][]{Turner2015}. It is
interesting to note that similar large temperature gradients were revealed
in a sample of molecular clouds located in the central zone of the Galaxy
by \cite{RodriguezFernandez2001} and \cite{Ao2013} who also suggested that
the dissipation of the supersonic turbulence could be responsible for large
molecular gas temperatures.

In the case of D1 cloud the model predicts large star formation efficiencies:
$\approx 87$\% in the case when parameter $\eta_{\star} = 1$ (that is probably
not consistent with the assumption of the residual gas retention) and
$\approx 63$\% when $\eta_{\star} = 5$. The last value is in agreement with the
large SFE in D1 cloud obtained by \cite{Turner2015}, the large SFE
($\sim 47\%$) in $\rho$ Oph cloud \citep{Wilking1983} and
agrees with the results of numerical simulations by \cite{Skinner2015} who
found that SFE may reach 50\% - 70\% in the case of the large gas opacity to
infrared radiation. It is also large enough to this cluster ends up as a
gravitationally bound super-star cluster \cite[see][]{Baumgardt2007,
Baumgardt2008}. It is important to note that $\eta_{\star}$ is not a
unique parameter that determines the value of the SFE. Model predictions
depend also on the star cluster mass and compactness. For example, the
SFE grew to 99\% in simulations with $\eta_{\star} = 5$ and core radius
$b = 0.1$pc. It is unlikely that at such large SFE the individual neighboring
winds and HII regions do not merge to disperse the parental cloud. We speculate
here that it is the star cluster compactness, that leads to a dramatic
difference between the deeply embedded into molecular cloud D1 cluster in
NGC 5253, and similar in mass and age, but gas-free cluster R136 in the 30 Dor
region (\cite{PortegiesZwart2010} and \cite{Mackey2003} estimated the R136
core radius to fall in the range 0.1pc - 0.3pc, see their Tables 3 and 4,
respectively). On the other hand, the model-predicted SFE drops rapidly when
one considers lower mass clusters. For example, in simulations with
$\eta_{\star} = 5$, $M_{SC} = 580$\Msol \, and $b = 0.14$pc \citep[parameters
similar to the Orion Nebula cluster (ONC),][]{Huff2006}, we obtained
$\approx 11$\% efficiency that agrees with the value of the SFE obtained by
\cite{Huff2006} for ONC and \cite{Megeath2016} for different stellar groups,
clusters and clouds in the Orion complex. Certainly, the above examples should
be considered only as an illustration because the stellar mass distribution in
these clusters differs from the Gaussian one (R136 and ONC were very well
fitted by different power-law profiles). In many other cases the star cluster
mass distribution is well represented by Moffat, Elson-Fall-Freeman, King
and Plummer models \citep[see][and references therein]{PortegiesZwart2010,
CuevasOtahola2020,Roeser2019}. We leave the discussion of different stellar
mass distributions to the future communication. 

It is instructive to note, that in spite of the large momentum input rate
allowed in the simulations (up to $5 \times L_{bol} / c$), the model-predicted
turbulence dissipation rate remains negligibly small in comparison with the
star cluster bolometric luminosity:
$2.2 \times 10^{-5} < L_{dis}/L_{bol} < 1.2 \times 10^{-4}$.
This favors the radiative feedback to be the major mechanism that supports
turbulence in this cloud, likely through the overpressurised HII regions formed
around most massive stars that move in the gravitational well of the cluster
\citep[e.g.][]{Mackey2013,Matzner2002,Krumholz2006}. Indeed, a low
radiative feedback efficiency is expected in dense, dusty environments
\citep[see][]{Haid2018}. This agrees with the fact that far infrared (FIR)
luminosity of the NGC 5253 central zone \citep[$\approx 8 \times 10^{42}
$erg s$^{-1}$][]{Cormier2015} is comparable and even exceeds the bolometric
luminosity of the D1 cluster. In the case of a low feedback efficiency
most of the deposited energy is radiated away in the IR regime, instead of
being used to unbind the residual gas. Therefore one must take care
comparing D1 cluster with Fig. 3 from \cite{Baumgardt2008} which confronts
the cloud binding to the accumulated radiative energy upon the assumption
of a 100\% radiative feedback efficiency.

Simulations with a Salpeter IMF with lower and upper cutoff masses 
3\Msol \, and 120\Msol \, result in slightly larger masses of the
residual gas ($\sim 3.9 \times 10^4$\Msol \, and $\sim 1.5 \times 10^5$\Msol)
and smaller star formation efficiencies ($\sim 80\%$ and $\sim 50\%$) in
models with $\eta_{\star} = 1$ and  $\eta_{\star} = 5$, respectively.

\section{Concluding remarks}

Here we studied the leftover gas density, velocity dispersion and temperature
in young stellar clusters with a given stellar mass distribution. It was
postulated that star formation in the collapsing cloud is altered via the
leftover gas stirring by massive stars which move in the gravitational well of
the cluster and the residual gas. It was also assumed that the cluster is
sufficiently compact and dense to prevent the leftover gas expulsion, and that
the post-star-forming system is stable. The last condition requires the
gradient of the turbulent pressure to be in balance with the gravitational pull
of the cluster and the rapidly dissipated turbulent energy to be regenerated
continuously. The last condition requires a sufficient number of massive
stars to be formed. Therefore the steady-state condition determines both, the
residual gas properties, and the star formation efficiency in clusters with a
suppressed star cluster wind.

We confront this model with properties of a compact cluster in a nearby dwarf
spheroidal galaxy NGC 5253 that is still deeply obscured by molecular cloud D1.
The model is in good agreement with several observed properties of this
cluster in spite that infalling  molecular filaments still supply gas to
the central zone of the galaxy \citep{Consiglio2017}. It predicts the
different stellar and gas density distributions with stellar mass more
concentrated towards the star cluster center.
The model-predicted mass-weighted velocity dispersion is in good agreement with
the observed value, while high molecular gas temperatures are consistent with
the large observed CO(3-2) over CO(2-1) intensity ratio. The large predicted 
star formation efficiency is sufficient for this cluster to end up as a
bound super star cluster.

The model suggests that turbulent energy dissipation may be an effective energy
source for the molecular gas heating in dense and compact strongly obscured
clusters as was also suggested by \cite{Pan2009}. It is likely that
turbulent heating results in molecular gas warm component that could be
detected in Far Infrared (FIR) low-excitation emission lines of oxygen, carbon
and other species and also in millimeter/submillimeter CO rotational lines,
while gas in Photon Dominated Regions (PDRs), directly heated by
Far Ultraviolet (FUV) and X-ray photons, is manifested by the
FIR high-excitation lines. CO and FIR emission lines in NGC 5253 were detected
by ALMA \citep[see][]{Turner2017} and Herschel \citep{Cormier2015}.
The observed [OI]63$\mu$m and [OI]145$\mu$m lines luminosity is about
$6.1 \times 10^{39}$erg s$^{-1}$ at the distance of about 4Mpc, that is larger
than the model-predicted turbulent energy dissipation rate. However the
Herschel observations do not separate D1 from other sources. The CO(3-2) line
luminosity is smaller, about $8.8 \times 10^{35}$erg s$^{-1}$. James Webb Space
Telescope(JWST) sensitivity and subarcsecond space resolution are required to
reveal the contribution of D1 cloud to the observed infrared emission. It is
also crucial to obtain better restrictions on the total CO luminosity and on
the temperature of warm molecular gas by observing higher J CO lines.

The fraction of the stellar feedback used to regenerate the turbulent energy
dissipation rate is not fixed in the present model and will be addressed in
a forthcoming communication. Nevertheless, the value of the momentum
input rate used in the simulations is motivated by the numerical
simulations and requires that only a tiny fraction of the radiation energy be
used to compensate the turbulent energy dissipation rate. 

One can apply this  model to clusters with an arbitrary mass distribution,
but must note that it is restricted to massive and compact clusters with a
suppressed mechanical feedback. It is likely, however, that our model
could be also applied to lower mass, less compact and massive, very compact
clusters with an extremely large SFE prior to their residual gas dispersal
if their parental clouds are supported against gravity by the turbulent
pressure and contract gradually in the quasi-static regime, as was suggested
by \cite{Huff2006} for the Orion Nebula cluster.

Certainly, the equilibrium conditions that we have used should change after
the onset of the supernova explosions. The further evolution of leftover gas
is beyond the scope of the present paper. We anticipate two
possible scenarios: the leftover gas could be expelled out of the cluster
by supernovae, that, however, is unlikely in systems with a sharp density
gradient \citep[see][]{Santiago2021}, or the turbulent energy dissipates after
the majority of massive stars explode as supernovae and the leftover gas,
enriched by massive star products, collapses to form a second stellar
generation.

\section*{Acknowledgements}

We thank our anonymous referee for his/her constructive comments and S. Beck
for careful reading of the manuscript. This study was supported by CONACYT,
M\'exico research grant A1-S-28458. JLT acknowledges the support of U.S.
National Science Foundation grant AST2006433. JM acknowledges support from a
Royal Society-Science Foundation Ireland University Research Fellowship and
S.M.G. acknowledges the support provided by CONACYT through C\'atedra n.482.

\bibliographystyle{apj2}
\bibliography{TUR}

\end{document}